# On the Inhibition of Linear Absorption in Opaque Materials Using Phase-Locked Harmonic Generation

Marco Centini[1], Vito Roppo[2], Eugenio Fazio[1], Federico Pettazzi[1], Concita Sibilia[1], Joseph W. Haus[3], John V. Foreman[4,5], Neset Akozbek[4], Mark J. Bloemer[4], Michael Scalora[4]

*1  Dipartimento di Energetica, University of Rome La Sapienza, Via Scarpa 16, Rome Italy*

*2  Departament de Fisica i Enginyeria Nuclear, Universitat Politecnica de Catalunya C/Colom 11, 08222 Terrassa Spain*

*3  Electro-Optics Program, University of Dayton, Dayton, OH 45469-0245, USA*

*4  Charles M. Bowden Research Facility, RDECOM, US Army Aviation and Missile Command, Redstone Arsenal, AL 35803*

*5 Department of Physics, Duke University, Durham, NC 27708*

## Abstract

We theoretically predict and experimentally demonstrate inhibition of linear absorption for phase and group velocity mismatched second and third harmonic generation in strongly absorbing materials, GaAs in particular, at frequencies above the absorption edge. A 100-fs pump pulse tuned to 1300nm generates 650nm and 435nm second and third harmonic pulses that propagate across a 450µm-thick GaAs substrate without being absorbed. We attribute this to a phase-locking mechanism that causes the pump to trap the harmonics and to impress them with its dispersive properties.

Since its inception in the early 1960s the study of nonlinear frequency conversion has focused on improving the efficiency of the process in transparent materials [1-6]. Nonlinear conversion rates depend on factors such as phase and group velocity mismatch, and peak pump intensity. Linear absorption is considered detrimental since it is assumed that the generated harmonics are reabsorbed. Harmonic generation in absorbing materials and/or semiconductors at frequencies above the absorption edge has been considered in the context of measuring the nonlinear coefficients [7-13], but it has not been as widely studied as in the transparency frequency range. For example, the enhancement of third-harmonic generation in various types of glasses opaque to third harmonic (TH) light was experimentally demonstrated [11]. UV second harmonic generation (SHG) above the absorption band edge in LiNbO3 [12] as well as UV and X-ray [13] SHG in semiconductors have been reported. These examples show that the subject is of interest partly for the purpose of realizing coherent sources, and because of the many potential applications that semiconductors find in optical technology.

Better understanding of propagation phenomena at or above the absorption edge of semiconductor and dielectric materials [14] is needed not only in view of the discrepancies that exist between predictions and experimental observations [11, 12], but also because in this range semiconductors like GaAs display a negative dielectric permittivity, thus raising the possibility of new effects related to negative refraction of light at optical and UV wavelengths, absorption notwithstanding. A systematic examination of dynamics and nonlinear frequency conversion above the absorption edge of semiconductors is still lacking primarily because these processes are thought to be uninteresting and inefficient, due to absorption and the naturally high degree of phase mismatch. In this Letter we dispel this notion, and predict and experimentally observe the inhibition of absorption for femtosecond, SH and TH (650nm and 435nm, respectively) generated signals in a GaAs substrate 450 microns thick. The characteristic absorption lengths of GaAs are much less than one micron. For example, transmittance through one micron of GaAs is $\sim 10^{-4}$ at 532nm, and $\sim 10^{-8}$ at 364nm. Our model also predicts much closer agreement with previous experimental observations of SH (266nm) transmission and reflection in a $LiNbO_3$ substrate compared to other approaches [12].

Bulk GaAs becomes opaque at wavelengths below 900nm. Our theory shows that the pump, tuned to a region of optical transparency, captures and impresses its dispersive properties on portions of the generated second and third harmonic signals, which in turn behave as parts of the pump and co-propagate for the entire length of the sample *without being absorbed*. This spectacular behavior is brought about by a phase locking phenomenon [15] that impacts harmonic generation and other types of parametric processes, with seeded or unseeded harmonic signals. These general conclusions thus apply: if the medium is transparent at the pump frequency, then the material will be transparent at the harmonic frequencies. Similarly, if the medium absorbs the pump and is transparent in the SH and TH ranges, the phase-locked components are absorbed and the normal components survive.

The prediction of a two-component SH signal (homogenous and inhomogeneous solutions of the wave equation) was made early on [4], and was later discussed and observed [5, 6]. In transparent materials the pulsed SHG process develops as follows: the inhomogeneous signal is trapped by the pump, and propagates at the pump's group velocity [15]. The homogeneous (normal) SH component propagates according to material dispersion at that frequency, and eventually walks off from the pump. In absorbing materials [7-13], Maker fringes [3] are observed as long as material absorption is small, while the amplitude of the transmitted beam is independent of sample thickness [12]. Thus, the evidence suggest that the pump interacts only with the normal SH component to produce Maker fringes. The interaction stops as soon as the normal component either is absorbed or walks off from the pump, leaving the inhomogeneous (phase locked) portion of the signal intact. Even though predictions and observations are not new, it was not until recently that in transparent materials these phenomena were cast in terms of a phase locking mechanism [15] that also impacts higher order nonlinearities [16]. Calculations show that the trapped signals acquire the dispersive properties of the pump, i.e. they propagate with the pump's index and group velocity. The theory tells us that if the group velocity mismatch is relatively large, the pump and the *normal* SH components separate immediately upon entering the medium, and the conversion process turns into a surface phenomenon: the pump field and its captive harmonic signals do not exchange energy inside the medium [15], i.e. their relative phase difference is constant, until another interface is crossed.

In absorbing materials, discrepancies have been recorded between predictions and experimental results [11, 12]. Even though there is recognition that the two SH components propagate at different group velocities and peculiar phase properties [17], a spectral analysis reveals a far more intimate connection than previously thought between the pump and the trapped harmonic pulses, i.e. a phase locking that binds the pulses spatially and temporally [15]. In the present analysis we calculate the energy velocities of the generated pulses, defined as usual: $V_e = \frac{<S>}{<U>}$; the brackets mean a definite integral. Experimental and theoretical evidence [5, 6, 15] shows that the pump and the phase locked pulses propagate at the same energy velocities. In confirmation of the fact that the generated fields behave in all respects as if they were pump fields, our calculations show that the proper energy velocities for the second and third harmonic fields (that is to say, the velocity of the pump field) are recovered only if the Landau energies [18], $U(z,t) = \frac{1}{8\pi}\left(\text{Re}\left(\frac{\partial[\omega\varepsilon(\omega)]}{\partial\omega}\right)|\mathbf{E}|^2 + \text{Re}\left(\frac{\partial[\omega\mu(\omega)]}{\partial\omega}\right)|\mathbf{H}|^2\right)$, are evaluated with the material parameters of the pump frequency. Therefore, we find that $V_e$ is identical for all pulses only if *the generated harmonic fields are attributed the material dispersion of the pump*. This finding cements the notion that the phase locked pulses behave as the pump pulse does, and ultimately require to be treated as such in the application of boundary conditions.

To model simultaneous second and third harmonic generation in ordinary materials we assume the fields may be decomposed as a superposition of harmonics:

$$\mathbf{E} = \hat{\mathbf{x}}\sum_{\ell=1}^{\infty}\left(E_{\ell\omega}(z,t)+c.c\right) = \hat{\mathbf{x}}\sum_{\ell=1}^{\infty}\left(\mathcal{E}_{\ell\omega}(z,t)e^{i\ell(kz-\omega t)}+c.c\right)$$
$$\mathbf{H} = \hat{\mathbf{y}}\sum_{\ell=1}^{\infty}\left(H_{\ell\omega}(z,t)+c.c\right) = \hat{\mathbf{y}}\sum_{\ell=1}^{\infty}\left(\mathcal{H}_{\ell\omega}(z,t)e^{i\ell(kz-\omega t)}+c.c\right) \qquad (1)$$

where $\mathcal{E}_{\ell\omega}, \mathcal{H}_{\ell\omega}$ are generic, spatially- and temporally-dependent, complex envelope functions; $k$ and $\omega$ are carrier wave vector and frequency, respectively, and $\ell$ is an integer. Eqs.(1) are a convenient representation of the fields, and no *a priori* assumptions are made about the envelopes. The linear response of the medium is described by a Lorentz oscillator model: $\varepsilon(\omega) = 1 - \frac{\omega_p^2}{\omega^2 + i\gamma\omega - \omega_r^2}$, and $\mu(\omega) = 1$, where $\gamma$, $\omega_p$, and $\omega_r$ are the damping coefficient, the plasma and resonant frequencies, respectively. Simultaneous second and third order electric nonlinearities are introduced as a nonlinear polarization of the type: $P_{NL} = \chi^{(2)}E^2 + \chi^{(3)}E^3$. Assuming that the polarization and currents may be decomposed in a manner similar to Eqs.(1), and that no diffraction is present, we obtain the following Maxwell-Lorentz system of Equations for the $\ell^{th}$ field components:

$$\frac{\partial \mathcal{E}_{\ell\omega}}{\partial \tau} = i\beta_{\ell\omega}\left(\mathcal{E}_{\ell\omega} - \mathcal{H}_{\ell\omega}\right) - 4\pi(\mathcal{J}_{\ell\omega} - i\beta_{\ell\omega}\mathcal{P}_{\ell\omega}) - \frac{\partial \mathcal{H}_{\ell\omega}}{\partial \xi} + i4\pi\beta_{\ell\omega}\mathcal{P}_{\ell\omega}^{NL} - 4\pi\frac{\partial \mathcal{P}_{\ell\omega}^{NL}}{\partial \tau}$$

$$\frac{\partial \mathcal{H}_{\ell\omega}}{\partial \tau} = i\beta_{\ell\omega}\left(\mathcal{H}_{\ell\omega} - \mathcal{E}_{\ell\omega}\right) - \frac{\partial \mathcal{E}_{\ell\omega}}{\partial \xi}$$

$$\frac{\partial \mathcal{J}_{\ell\omega}}{\partial \tau} = \left(2i\beta_{\ell\omega} - \gamma_{\ell\omega}\right)\mathcal{J}_{\ell\omega} + \left(\beta_{\ell\omega}^2 + i\gamma\beta_{\ell\omega} - \beta_{r,\ell\omega}^2\right)\mathcal{P}_{\ell\omega} + \pi\omega_{p,\ell\omega}^2 \mathcal{E}_{\ell\omega}$$

$$\frac{\partial \mathcal{P}_{\ell\omega}}{\partial \tau} = \mathcal{J}_{\ell\omega}$$

(2)

where $\mathcal{J}_{\ell\omega}$, $\mathcal{P}_{\ell\omega}$, $\mathcal{P}_{\ell\omega}^{NL}$ are the linear current, the linear and nonlinear polarizations, respectively. The coordinates are scaled so that $\xi = z/\lambda_0$, $\tau = ct/\lambda_0$, $\omega_0 = \frac{2\pi c}{\lambda_0}$, where $\lambda_0 = 1\mu m$ is a convenient reference wavelength; $\gamma_{\ell\omega}$, $\beta_{\ell\omega} = 2\pi\ell\omega/\omega_0$, $\beta_{r,\ell\omega} = 2\pi\omega_{r,\ell\omega}/\omega_0$, and $\omega_{p,\ell\omega}$ are the scaled damping coefficient, resonance and plasma frequencies for the $\ell^{th}$ harmonic, respectively. Expanding the field powers in terms of generic envelope functions leads to:

$$\mathcal{P}_{\omega}^{NL} = 2\chi^{(2)}\left(\mathcal{E}_{2\omega}^* \mathcal{E}_{3\omega} + \mathcal{E}_{\omega}^* \mathcal{E}_{2\omega}\right)$$
$$+ 3\chi^{(3)}\left(|\mathcal{E}_{\omega}|^2 \mathcal{E}_{\omega} + \mathcal{E}_{2\omega}^2 \mathcal{E}_{3\omega}^* + \mathcal{E}_{3\omega}\mathcal{E}_{\omega}^{*2} + 2|\mathcal{E}_{2\omega}|^2 \mathcal{E}_{\omega} + 2|\mathcal{E}_{3\omega}|^2 \mathcal{E}_{\omega}\right)$$
$$\mathcal{P}_{2\omega}^{NL} = \chi^{(2)}\left(\mathcal{E}_{\omega}^2 + 2\mathcal{E}_{\omega}^* \mathcal{E}_{3\omega}\right) + 3\chi^{(3)}\left(|\mathcal{E}_{2\omega}|^2 \mathcal{E}_{2\omega} + 2|\mathcal{E}_{3\omega}|^2 \mathcal{E}_{2\omega} + 2\mathcal{E}_{2\omega}^* \mathcal{E}_{3\omega}\mathcal{E}_{\omega} + 2|\mathcal{E}_{\omega}|^2 \mathcal{E}_{2\omega}\right)$$
$$\mathcal{P}_{3\omega}^{NL} = 2\chi^{(2)}\mathcal{E}_{2\omega}\mathcal{E}_{\omega} + \chi^{(3)}\left(\mathcal{E}_{\omega}^3 + 6|\mathcal{E}_{2\omega}|^2 \mathcal{E}_{3\omega} + 3|\mathcal{E}_{3\omega}|^2 \mathcal{E}_{3\omega} + 3\mathcal{E}_{2\omega}^2 \mathcal{E}_{\omega}^* + 6|\mathcal{E}_{\omega}|^2 \mathcal{E}_{3\omega}\right)$$

(3).

Eqs.(2) represent twelve nonlinear coupled equations supplemented by Eqs.(3), and are solved using a time-domain, split-step, fast Fourier transform-based pulse propagation algorithm [19].

In Fig.(1) we illustrate the process in a generic, dispersive material for an incident 100fs pump pulse tuned to 900nm. The oscillator parameters are found in the caption. We assume the fields are tuned far enough from resonance to render absorption negligible. The incident peak pump power is chosen to be just a few Watts/cm$^2$, $\chi^{(2)} \sim 10^{-7} \, esu \, (\sim 1 pm/V)$, and $\chi^{(3)} \sim 10^{-12} \, esu$ (n$_2$~10$^{-15}$ cm$^2$/W). Figs.(1) show that both normal and trapped components. The normal components travel at a slower group velocity than the pump and eventually walk off.

The pump and the trapped portions of the SH and TH signals travel at the same energy velocity. In Fig.2 we depict the k-space power spectrum of the signals inside the material. We find two-component SH and TH signals, one phase locked, having wave-vectors twice and three times that of pump, the other normal, having wave-vectors as predicted by material dispersion. The generation of the normal and phase-locked components is thus a fundamental property of the nonlinear process, independent of material parameters [15].

What happens when significant absorption is present? We consider propagation inside a 450-micron thick GaAs substrate, and tune a ~160MW/cm$^2$ incident pump pulse to 1300nm. The peak power is low enough to avoid shape changes due to self- and cross-phase modulation, and nonlinear pump absorption. Both SH and TH signals are tuned well-above the absorption edge of GaAs, at 650nm and 435nm, respectively. Allowing for differences in the index of refraction of GaAs [14] relative to the generic material of Fig.1, we simply state that the results for GaAs are identical to those displayed in Figs.(1-2), except that now the normal components are absent in both figures, and the ordinate is scaled differently to reflect the higher incident peak powers. Our calculations thus prove that the introduction of absorption causes the trapped SH and TH signals to survive in the form of phase locked pulses, and to propagate undisturbed across the substrate; the respective normal components are absorbed within a fraction of a micron from the entry surface of the sample. A spectral analysis (k and ω) of the fields confirms that phase locking and energy clamping (no energy is exchanged) occur between the pump and the harmonic fields: the phase locking prevents the pump from recognizing the harmonic fields as such, since all the fields behave with the dispersive properties of the pump.

The survival of the trapped SH and TH fields was experimentally verified using the setup shown in Fig.3. An optical parametric amplifier pumped by ~100 fs, amplified Ti:Sapphire pulses generated the pump at 1300nm. The collimated pump irradiated the GaAs substrate at an angle of 20°. After passing through the GaAs, the SH and TH fields were separated from the pump using a prism and lens/slit assembly. The signals were coupled to a spectrometer by means of a liquid light guide and were then measured using a near infrared photomultiplier tube (for the pump) or liquid nitrogen cooled CCD array (for SH and TH). The measured spectra for the pump and its first two harmonics are shown in Fig.4. Absolute values of the SH and TH energies were calibrated by sending laser beams of known power along the same optical paths and through the same detection system, and we were able to estimate the SH and TH conversion efficiencies to be $1.3\times10^{-8}$ and $2.5\times10^{-9}$, respectively. Removing the GaAs sample from the beam path caused the conversion efficiencies to drop by a factor of twenty for the SH, and by a factor of two for the TH. Conversion efficiencies and pump transmittance data were used in the model to estimate $\chi^{(2)}\sim100$pm/V, and $n_2\sim10^{-12}$ cm$^2$/W, respectively. Finally, we comment on the results reported in reference [12]. A LiNbO$_3$ sample was pumped at 532nm to produce transmitted and reflected signals at 266nm (absorption edge at 290nm). The ratio of transmitted to reflected SH intensities was measured to be 23±4, and the predicted value was 40. The discrepancy was attributed to the fact that the material was illuminated at an angle of 10°, while the calculation was done at normal incidence. In comparison, our model predicts a transmitted/reflected SH signal ratio of ~20 at normal incidence, much closer to the reported measurement. Thus Eqs.(2) correctly accounts for any dynamical index and group velocity modifications of the harmonic signals, and accurately describe events unfolding inside the sample and at all interfaces.

In conclusion, we have presented theoretical and experimental evidence that absorption can be inhibited in opaque materials, well above the absorption edge of semiconductors and dielectrics alike. This dramatic result is due to a phase locking mechanisms that causes the pump to trap and impress its dispersive properties to the generated signals. We have shown that a GaAs substrate 450 microns thick supports the propagation of red and violet light. We have also obtained good agreement with previous experimental results that reported SH generation from LiNbO$_3$ at 266nm. Our results thus suggest that it is possible to achieve relatively efficient nonlinear frequency conversion at high frequencies, particularly towards the UV, using readily available sources and materials.

**Figure Captions**

**Fig.1:** Pump (900nm), second (450nm) and third (300nm) harmonic pulses propagating inside a generic material. The Lorenz parameters are: $\omega_p=9$, $\omega_r=4.7$, $\gamma=0$. Both phase locked and normal components are shown, the first trapped by the pump, the second lagging behind due to walk-off.

**Fig.2:** k-space spectrum of the fields depicted in Fig.1. The trapped pulses carry most of the energy contained in the harmonic signals.

**Fig.3:** Experimental setup. A beam generated by an optical parametric amplifier (OPA) irradiates a GaAs substrate. After passing through the sample, a prism (EDP) separates the harmonics. The three beams are coupled to a spectrometer via a liquid light guide (LLG) and measured using a near infrared PMT (for the pump) or a CCD (for the SH and TH). A moveable slit assembly provides spatial filtering so that only one beam reaches the LLG.

**Fig.4:** Measured spectra of the pump and the transmitted harmonics. The area under each spectrum corresponds to the measured energy (2.9 µJ for pump, 3.7x10$^{-8}$ µJ for SH, 7.3x10$^{-9}$ µJ for TH).

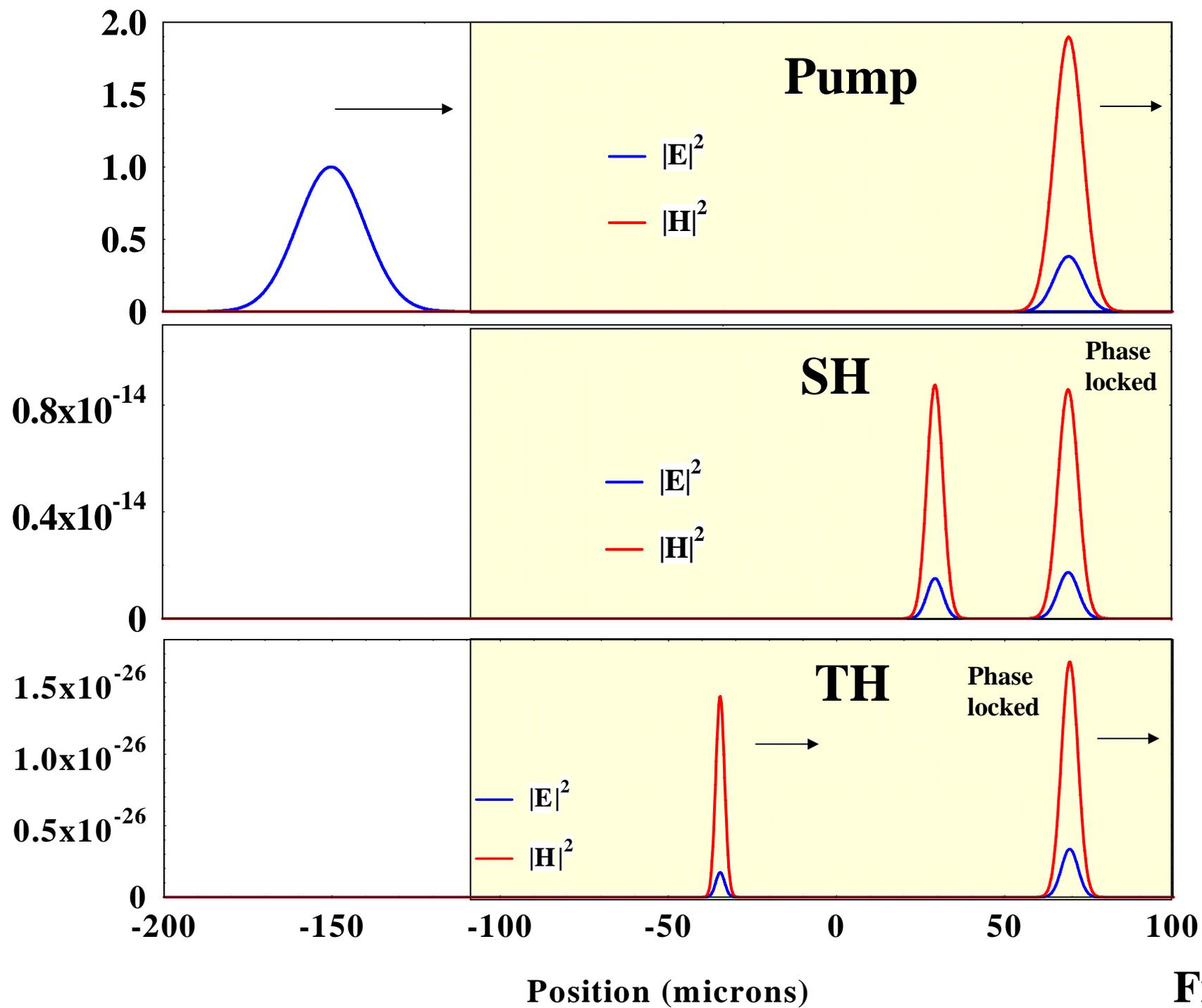

Fig.1

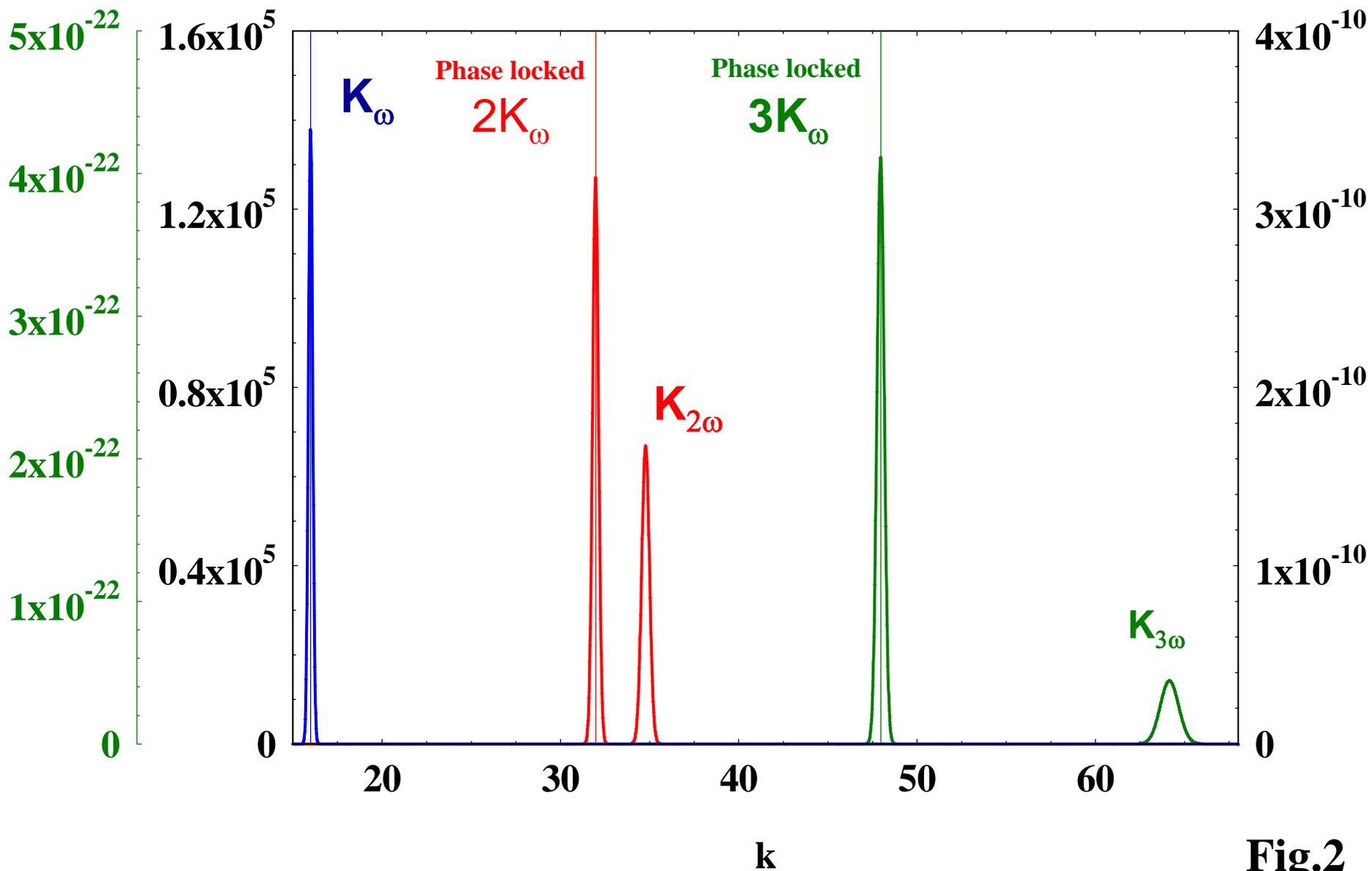

Fig.2

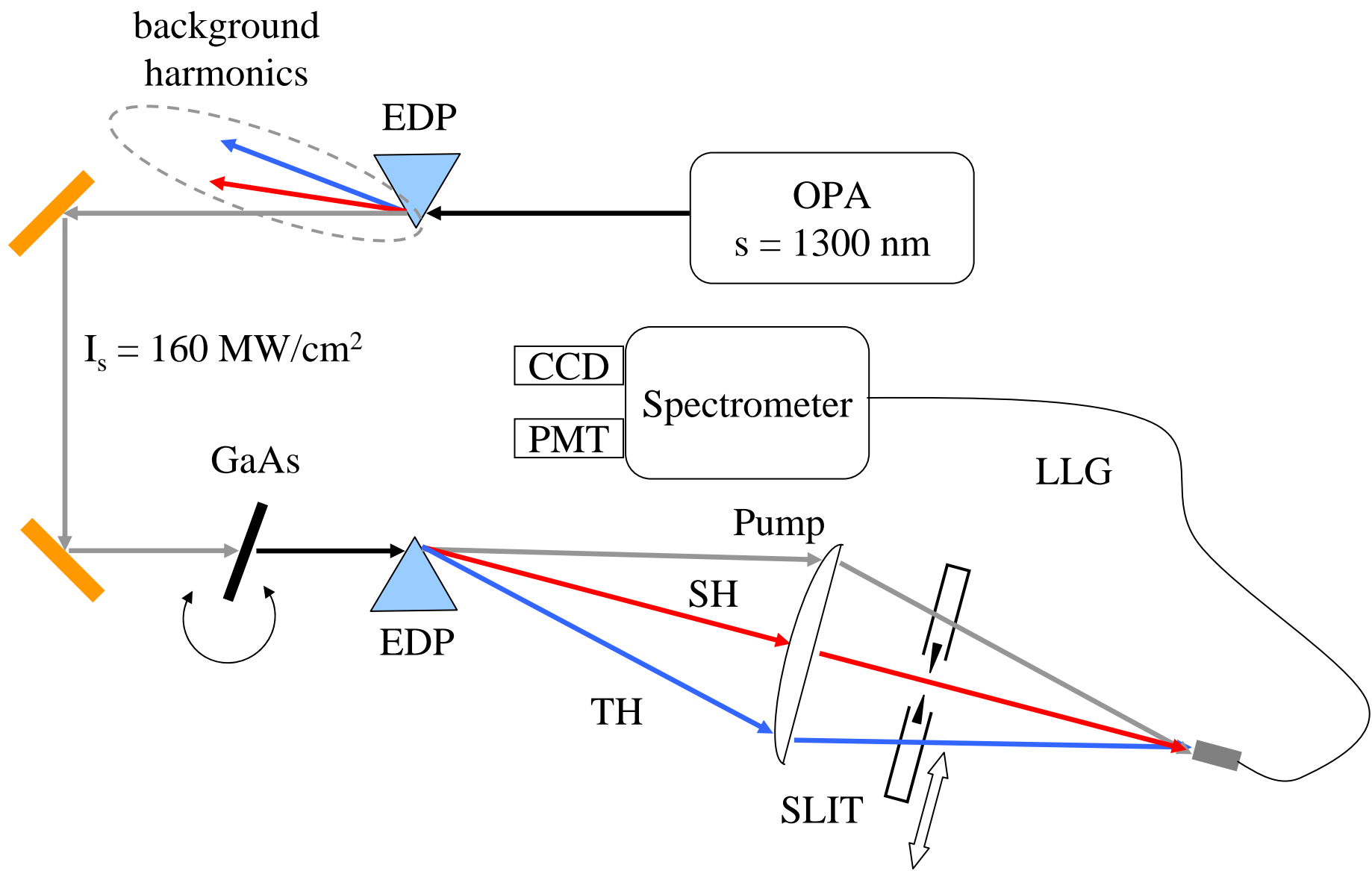

Fig.3

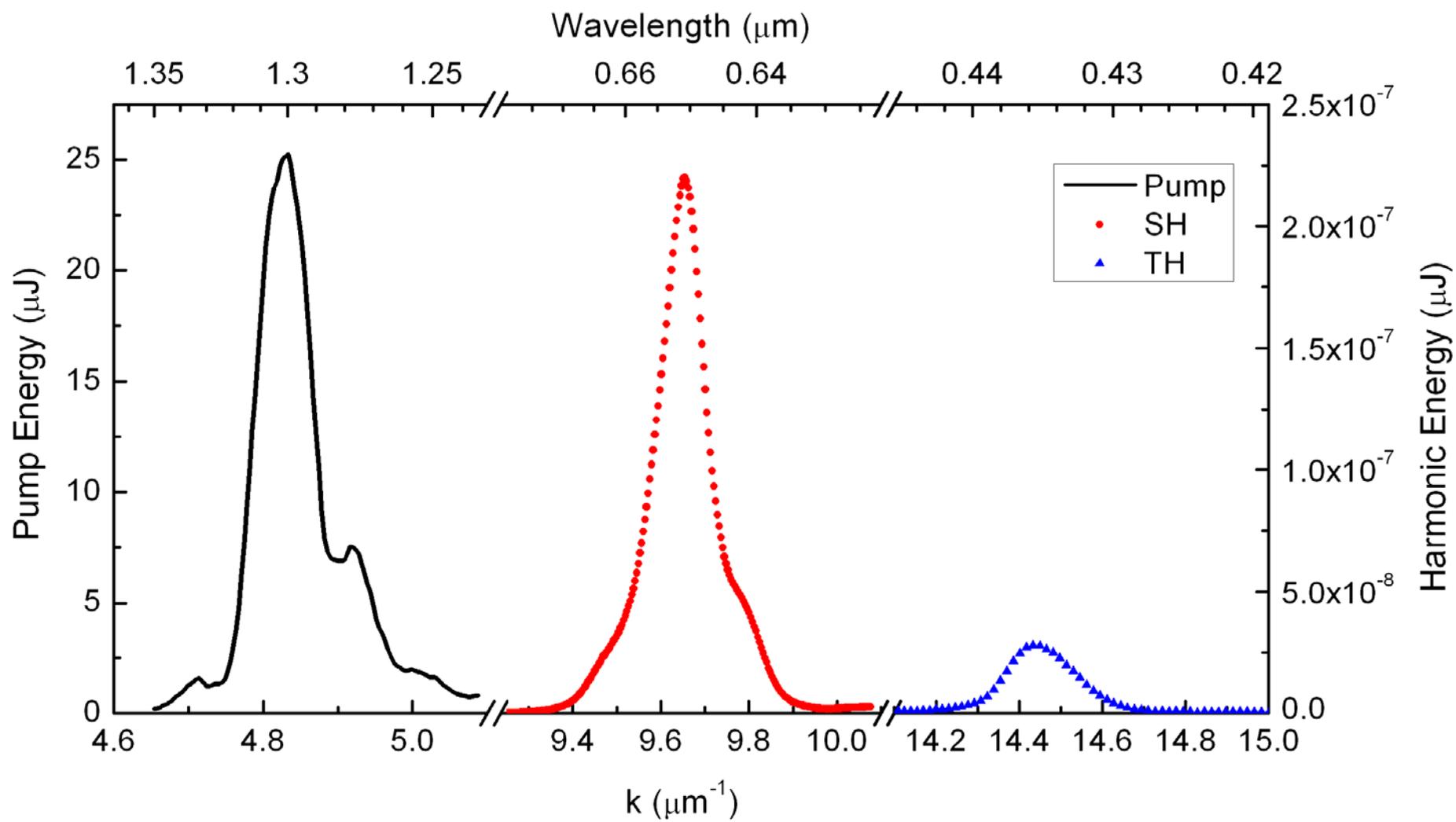

Fig.4